\documentclass[
    ,final            
  ]
  {aipproc}

\layoutstyle{6x9}

\newcommand{\beq}{\begin{equation}}
\newcommand{\eeq}{\end{equation}}
\newcommand{\be}{\begin{eqnarray}}
\newcommand{\ee}{\end{eqnarray}}

\def\gsim{\buildrel > \over {_{\sim}}}

\begin{document}

\title{Spectral Functions and Nuclear Response}


\author{Omar Benhar}{
  address={INFN, Sezione di Roma \\ 
Dipartimento di Fisica, Universit\`a ``La Sapienza'' \\
I-00185 Roma, Italy}
}

\begin{abstract}
I discuss the relation between the nuclear response and the 
Green function describing the propagation of a nucleon in 
the nuclear medium. Within this formalism, the widely used 
expressions in terms of spectral functions can be derived 
in a consistent and rigorous fashion. The results of recent 
applications to the study of the inclusive electron-nucleus 
cross section in the impulse approximation regime are brielfy 
analyzed.

\end{abstract}

\maketitle


\section{Introduction}

Within non relativistic many-body theory, the nuclear response to a scalar probe delivering 
momentum {\bf q} and energy $\omega$ can be written in terms of the the imaginary part 
of the polarization propagator $\Pi({\bf q},\omega)$ according to 
\cite{FetterWalecka,BFF:1992}
\beq
\label{def:resp}
S({\bf q},\omega) = \frac{1}{\pi}\ {\rm Im}\  \Pi({\bf q},\omega) = \frac{1}{\pi}\ 
{\rm Im}\ \langle 0 \vert \rho^\dagger_{{\bf q}}  \
\frac{1}{H-E_0-\omega-i\eta} \ \rho_{{\bf q}} \vert 0 \rangle \ ,
\eeq
where $\eta=0^+$, $\rho_{{\bf q}}= \sum_{{\bf k}} a^\dagger_{{\bf k}+{\bf q}} a_{{\bf k}}$ 
is the operator describing the fluctuation of the target density induced by the interaction 
with the probe, 
$a^\dagger_{{\bf k}}$ and $a_{{\bf k}}$ are nucleon creation
and annihilation operators, $H$ is the nuclear hamiltonian and $\vert 0 \rangle$ is the 
target ground state, satisfying the Schr\"odinger equation 
$H\vert 0 \rangle = E_0 \vert 0 \rangle$.

In this short note, I will discuss the relation between $S({\bf q},\omega)$ and the nucleon 
Green function, leading to the popular expression of the response in terms of nucleon 
spectral functions \cite{BFF:1992,BFF:1989}. The main purpose of this work is 
show that the spectral function formalism, while being often advocated using heuristic 
arguments, can be derived in a rigorous and fully consistent fashion.

For the sake of simplicity, I will consider uniform nuclear matter with
equal numbers of protons and neutrons. In the Fermi gas (FG) model, i.e. neglecting
all interactions, such a system reduces to a degenerate Fermi gas of density
$\rho = A/V$, $A$ and $V$ being the number of nucleons and the normalization volume,
respectively. In the FG ground state the $A$ nucleons occupy all momentum eigenstates
belonging to the eigenvalues ${\bf k}$ such that $|{\bf k}| < k_F$, 
$k_F = 2 \rho/3 \pi^2$ being the Fermi momentum.
 
\section{Formalism}

Equation (\ref{def:resp}) clearly shows that the interaction with the probe leads to
a transition of the struck nucleon from a {\em hole state} of momentum ${\bf k}$
to a {\em particle state} of momentum 
${\bf k}+{\bf q}$.
To obtain $S({\bf q},\omega)$ one needs to describe the propagation of the resulting
particle-hole pair through the nuclear medium.

The fundamental quantity involved in the theoretical treatment of many-body system is 
the Green function, i.e. the quantum mechanical amplitude associated with the propagation
of a particle from $x\equiv(t,{\bf x})$ to $x^\prime\equiv(t^\prime,{\bf x}^\prime)$
\cite{FetterWalecka}. In nuclear matter, due to translation invariance, 
the Green function only depends on the difference $x-x^\prime$, and after Fourier 
transformation to the conjugate variable $k\equiv({\bf k},E)$ can be written in the form
\be
\nonumber
G({\bf k},E) & = & \langle 0 \vert a^\dagger_{{\bf k}} \ 
\frac{1}{H-E_0-E-i\eta} \ a_{{\bf k}} \vert 0 \rangle 
 - \langle 0 \vert a_{{\bf k}} \ \frac{1}{H-E_0+E-i\eta} \ a^\dagger_{{\bf k}} \vert 0 \rangle \\
 & = &  G_h({\bf k},E) + G_p({\bf k},E) \ ,
\label{green:1}
\ee
where $G_h$ and $G_p$ correspond to propagation of nucleons sitting in hole and 
particle states, respectively. 


The connection between Green function and spectral functions is established
through the Lehman representation \cite{FetterWalecka}
\beq
G({\bf k},E) = \int dE^\prime \left[ \frac{P_h({\bf k},E^\prime)}{E^\prime - E - i\eta}
    -  \frac{P_p({\bf k},E^\prime)}{E - E^\prime - i\eta} \right] \ , 
\eeq
implying 
\be
P_h({\bf k},E) & = & \sum_n 
\vert \langle n_{(N-1)}(-{\bf k}) \vert a_{{\bf k}} \vert 0_N \rangle \vert^2
\delta(E-E^{(-)}_n+E_0) = \frac{1}{\pi}\ {\rm Im}\ G_h({\bf k},E) \ , \\
\label{def:Ph}
P_p({\bf k},E) & = & \sum_n 
\vert \langle n_{(N+1)}({\bf k}) \vert a^\dagger_{{\bf k}} \vert 0_N \rangle \vert^2
\delta(E+E^{(+)}_n-E_0) = \frac{1}{\pi} {\rm Im}\ G_p({\bf p},E) \ ,
\label{def:Pp}
\ee
where $\vert \langle n_{(N \pm 1)}(\pm {\bf k}) \rangle$ denotes an eigenstate of the 
$(A \pm 1)$-nucleon system, carrying momentum $\pm{\bf k}$ and energy $E^{(\pm)}_n$.

Within the FG model the matrix elements of the creation and annihilation operators
reduce to step functions, and the Green function takes a very simple form. 
For example, for hole states we find
\beq
G_{FG,h}({\bf k},E) = \frac{\theta(k_F-|{\bf k}|)}{E+\epsilon^0_k-i\eta}  \ ,
\label{green:FG}
\eeq
with $\epsilon^0_k = |{\bf k}^2| /2M$, $M$ being the nucleon mass, implying 
\beq
P_{FG,h}({\bf k},E) = \theta(k_F-|{\bf k}|) \delta(E+\epsilon^0_k) \ .
\eeq

Strong interactions modify the energy of a nucleon carrying momentum ${\bf k}$ according to
$\epsilon^0_k \longrightarrow \epsilon^0_k + \Sigma({\bf k},E)$, where $\Sigma({\bf k},E)$
is the {\em complex} nucleon self-energy, describing the effect of nuclear dynamics. 
As a consequence, the Green function for hole states becomes 
\beq
G_h({\bf k},E) = \frac{1}{ E + \epsilon^0_k - \Sigma({\bf k},E)} \ .
\label{greenh:2}
\eeq

A very convenient decomposition of $G_h({\bf k},E)$ 
can be obtained inserting a complete set of $(A-1)$-nucleon states  
(see Eqs.(\ref{green:1})-(\ref{def:Ph})) and isolating the contributions of
one-hole {\em bound} states, whose weight is given by \cite{BFF:1990}
\beq
Z_k = | \langle -{\bf k} | a_{{\bf k}} | 0 \rangle |^2 = \theta(k_F-|{\bf k}|) 
\Phi_k \ .
\label{def:Z}
\eeq
Note that in the FG model these are the only nonvanishing terms, 
and $\Phi_k \equiv 1$, while in the presence of interactions
$\Phi_k < 1$.
The resulting contribution to the Green function exhibits a pole at 
$-\epsilon_k$, the {\em quasiparticle} energy $\epsilon_k$ being defined 
through the equation
\beq
\epsilon_k = \epsilon^0_k + {\rm Re}\ \Sigma({\bf k},\epsilon_k) \ .
\label{QP:energy}
\eeq
The full Green function can be rewritten 
\beq
G_h({\bf k},E) = \frac{Z_k}{E+\epsilon_p+i Z_k\ {\rm Im}\ \Sigma({\bf k},e_k)}
 + G^B_h({\bf k},E) \ ,
\eeq
where $G^B_h$ is a smooth contribution, asociated with $(A-1)$-nucleon states 
having at least one nucleon excited to the continuum (two hole-one particle, 
three hole-two particles \ldots) due to virtual scattering processes
induced by nucleon-nucleon (NN) interactions. The corresponding spectral function is 
\beq
P_h({\bf k},E) =  \frac{1}{\pi}\
\frac{ Z_k^2 \ {\rm Im}\ \Sigma({\bf k},\epsilon_k) }
{ [E + \epsilon^0_k + {\rm Re}\ \Sigma({\bf k},\epsilon_k)]^2 +
          [Z_k {\rm Im}\ \Sigma({\bf k},\epsilon_k)]^2 }
 + P^B_h({\bf k},E) \ .
\eeq
The first term in the right hand side of the above equation yields the spectrum of
a system of independent quasiparticles, carrying momenta $|{\bf k}|<k_F$, moving in 
a complex mean field whose real and imaginary parts determine the quasiparticle 
effective mass and lifetime, respectively. The presence of the second term is 
a consequence of nucleon-nucleon correlations, not taken into account in the mean 
field picture. Being the only one surviving at $|{\bf k}|>k_F$, in the FG model 
this correlation term vanishes.

\begin{figure}[ht]
  \includegraphics[height=.30\textheight]{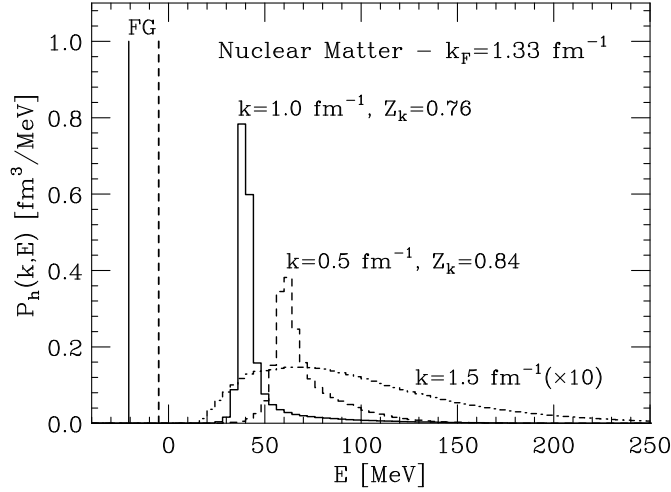}
  \caption{Energy dependence of the hole spectral function of nuclear matter
\cite{BFF:1989}.
The solid, dashed and dot-dash lines correspond to $|{\bf k}|=$ 1, 0.5 and 1.5 fm$^{-1}$,
respectively. The FG spectral function at $|{\bf k}|=$ 1 and 0.5 fm$^{-1}$ is shown
for comparison. 
The quasiparticle strengths of Eq.(\ref{def:Z}), are also reported. \label{fig1} }
\end{figure}

Figure \ref{fig1} illustrates the energy dependence of the hole spectral function
of nuclear matter, calculated in Ref. \cite{BFF:1989} with a realistic nuclear 
hamiltonian yielding an accurate description of NN scattering data up to pion production 
threshold. Comparison with the FG model clearly shows
that the effects of nuclear dynamics and NN correlations are large, resulting 
in a shift of the quasiparticle peaks, whose finite width becomes 
large for deeply-bound states with $|{\bf k}| \ll k_F$. In addition, NN 
correlations are responsible 
for the appearance of strength at $|{\bf k}|>k_F$. The energy integral
\beq
\label{def:nk}
n(k) = \int dE\  P_h({\bf k},E)
\eeq
yields the occupation probability of the state of momentum ${\bf k}$. The results 
of Fig. \ref{fig1} clearly show that in presence of correlations 
$n(|{\bf k}|>k_F)\neq0$.

In general, the calculation of the response requires the knowledge of $P_h$ and $P_p$, as 
well as of the particle-hole effective interaction \cite{BFF:1992,WIM:2004}.
The spectral functions are mostly affected by short range NN correlations 
(see Fig. \ref{fig1}), while the inclusion of the effective interaction, e.g. within the 
framework of the Random Phase Approximation (RPA) \cite{WIM:2004}, is needed to account
for collective excitations induced by long range correlations, involving more than 
two nucleons. 

At large momentum transfer, as the space resolution of the probe becomes small compared 
to the average NN separation distance, $S({\bf q},\omega)$ is no longer significantly 
affected by long range correlations. In this kinematical regime the zero-th order 
approximation in the effective interaction is expected to be applicable, and the 
response can be written in the simple form
\beq
\label{L0}
S({\bf q},\omega) = \int d^3k dE\ P_h({\bf k},E) P_p({\bf k}+{\bf q},\omega-E) \ .
\eeq
The widely employed impulse approximation (IA) can be readily obtained from 
the above definition replacing $P_p$ with the FG result, which amounts to 
disregarding final state interactions (FSI) betwen the struck nucleon and the spectator
particles: 
\beq
\label{IA}
S_{IA}({\bf q},\omega) = \int d^3k dE\ P_h({\bf k},E) \theta(|{\bf k}+{\bf q}|-k_F)
\delta(\omega-E-\epsilon^0_{|{\bf k}+{\bf q}|}) \ .
\eeq

At moderate momentum transfer, both the full response and the particle and hole
spectral functions can be obtained using non relativistic many-body theory.
The results of Ref.\cite{BFF:1992} suggest that the zero-th order approximations
of Eqs.(\ref{L0}) and (\ref{IA}) are fairly accurate at $|{\bf q}|~\gsim~500$~MeV.
However, it has to be pointed out that in this kinematical 
regime the motion of the struck nucleon in the final state can no longer 
be described using the non relativistic formalism.
While at IA level this problem can be easily circumvented, replacing the 
non relativistic kinetic energy with its relativistic counterpart, 
obtaining the response at large $|{\bf q}|$ from Eq.(\ref{L0}) involves
further approximations, needed to calculate of the particle spectral 
function.

A systematic scheme to include corrections to Eq.(\ref{IA}) and take into 
account FSI effects, originally proposed in Ref. \cite{gangofsix}, is discussed
in Ref. \cite{marianthi}. In the simplest implementation of this approach the 
reponse is obtained from the IA result according to
\beq
S({\bf q},\omega) = \int d\omega^\prime\ S_{IA}({\bf q},\omega^\prime) 
 F_{{\bf q}}(\omega-\omega^\prime) \ ,
\label{S:fold}
\eeq
the folding function $F_{{\bf q}}$ being related to the particle spectral function
through
\beq
F_{{\bf q}}(\omega-E-\epsilon^0_{{\bf q}}) =
P_p({\bf q},\omega-E)  \ ,
\label{f:fold}
\eeq
with $\epsilon^0_{{\bf q}} = \sqrt{{\bf q}^2+M^2}$.
Obvioulsy, at large ${\bf q}$ the calculation of $P_p({\bf q},\omega-E) $ cannot 
be carried out using a nuclear potential model. Hovever, $F_{{\bf q}}$ can be 
obtained form the measured NN scattering amplitude within the eikonal 
approximation \cite{marianthi}. It has to be pointed out that NN correlation, whose 
effect on $P_h$ is illustrated in Fig. \ref{fig1}, also 
affect the particle spectral function and, as a consequence, the folding function 
of Eq. (\ref{f:fold}). In the absence of FSI $F_{{\bf q}}$ shrinks to 
a $\delta$-function and the IA result of Eq.(\ref{IA}) is recovered.

\section{Applications to lepton-nucleus scattering}

The formalism outlined in the previous section can be readily generalized to describe
lepton-nucleus scattering, replacing the density fluctuation operator $\rho_{{\bf q}}$
with the appropriate vector and axial-vector currents.
The large body of theoretical and experimental work on inclusive electron-nucleus 
scattering has been recently reviewed in Ref. \cite{RMP}.

Over the past few years, significant effort has been devoted to the study of 
the kinematical region 
corresponding to beam energies around 1 GeV, whose understanding is relevant to 
the analysis of many neutrino oscillation experiments \cite{PRD}. 

\begin{figure}[ht]
  \includegraphics[height=.40\textheight]{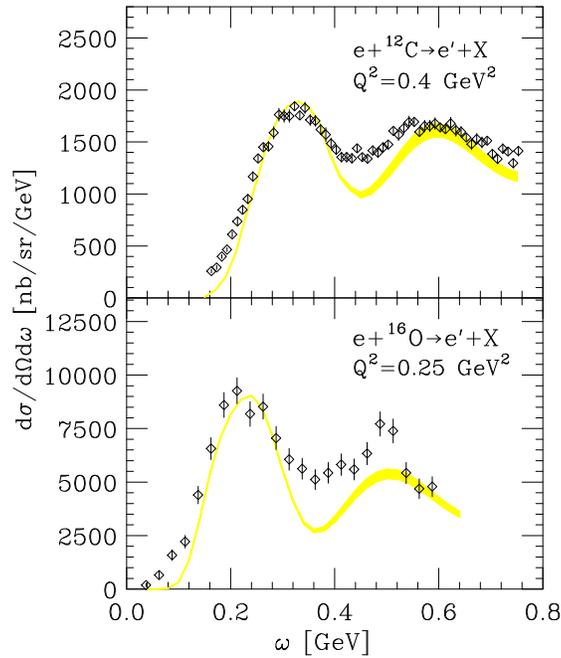}
  \caption{ \small
Upper panel: inclusive electron scattering cross section off carbon at
beam energy 1.3 GeV and scattering angle 37.5$^\circ$, as a function of the 
electron energy loss $\omega$. The shaded area shows the results of 
Ref. \cite{Benhar06}.
Data from Ref. \cite{Sealock89}. Lower panel: same as in the
upper panel, but for oxygen target, beam energy 1.2 GeV and scattering 
angle 32$^\circ$. Data from Ref. \cite{Anghinolfi96}. \label{fig2} }
\end{figure}

In Fig. \ref{fig2} the results of Ref. \cite{Benhar06}, obtained using 
the realistic hole spectral functions of Ref. \cite{LDA} and the particle spectral functions 
resulting from the approach of Ref. \cite{gangofsix}, are compared to 
the measured electron scattering cross sections off Carbon and Oxygen
of Refs. \cite{Sealock89} and \cite{Anghinolfi96}, respectively.
It appears that, while for the kinematics corresponding to the higher value of 
$Q^2 = |{\bf q}|^2-\omega^2$ the peaks corresponding to quasi-elastic scattering 
and delta resonance production are both very well described, at the lower $Q^2$
the delta peak is somewhat underestimated. In both cases, a sizable deficit of strength 
is observed in the region of the dip betwen the two peaks. The possibility that these
problems may be ascribed to deficiencies in the description of the elementary 
electron-nucleon cross section above pion production threshold is being actively 
investigated \cite{NUINT07}.

\begin{theacknowledgments}
This paper is dedicated to the memory of Adelchi Fabrocini and Vijay Pandharipande,
whose work led to important and lasting progress in nuclear 
response theory.
\end{theacknowledgments}



\bibliographystyle{aipproc}   

\bibliography{benhar}

\IfFileExists{\jobname.bbl}{}
 {\typeout{}
  \typeout{******************************************}
  \typeout{** Please run "bibtex \jobname" to optain}
  \typeout{** the bibliography and then re-run LaTeX}
  \typeout{** twice to fix the references!}
  \typeout{******************************************}
  \typeout{}
 }

\end{document}